%% file: prl_v7.tex
\documentclass[prd,nofootinbib,preprintnumbers,twocolumn,showpacs,floatfix,superscriptaddress]{revtex4-1}

\usepackage[T1]{fontenc}
\usepackage[utf8]{inputenc}
\usepackage{subfigure, lmodern, amsmath,amssymb, graphicx, pifont, adjustbox, bm, xcolor}
\usepackage{amsfonts}
\usepackage{comment}
\usepackage{mathtools}
\usepackage{float}
\usepackage{ braket }
\usepackage{longtable}

\renewcommand{\upuparrows}{\mathbin\uparrow\hspace{-.3em}\uparrow}

\DeclarePairedDelimiter{\ceil}{\lceil}{\rceil}
\DeclarePairedDelimiter\floor{\lfloor}{\rfloor}
\makeatletter\g@addto@macro\bfseries{\boldmath}\makeatother

\usepackage[colorlinks=true, citecolor=blue, urlcolor=blue, linkcolor=blue, breaklinks=true, pdfpagelabels=false]{hyperref}

\newcommand{\appendixref}[1]{\hyperref[#1]{appendix~\ref{#1}}}
\def\equationautorefname~#1\null{eq.\,(#1)\null}

\input{shortcuts}

\begin{document}

\title{The silence of binary Kerr}

\author{Rafael Aoude}
\affiliation{PRISMA$^+$  Cluster of Excellence \& Mainz Institute for Theoretical Physics, Johannes Gutenberg-Universit\"at Mainz, 55099 Mainz, Germany} 
\author{Ming-Zhi Chung}
\affiliation{Department of Physics and Astronomy, National Taiwan University, Taipei 10617, Taiwan}
\author{Yu-tin Huang}
\affiliation{Department of Physics and Astronomy, National Taiwan University, Taipei 10617, Taiwan}
\affiliation{Physics Division, National Center for Theoretical Sciences, National Tsing-Hua University, No.101,
Section 2, Kuang-Fu Road, Hsinchu, Taiwan}
\author{Camila S.\ Machado}
\affiliation{PRISMA$^+$  Cluster of Excellence \& Mainz Institute for Theoretical Physics, Johannes Gutenberg-Universit\"at Mainz, 55099 Mainz, Germany} 
\author{Man-Kuan Tam}
\affiliation{Department of Physics and Astronomy, National Taiwan University, Taipei 10617, Taiwan}
\preprint{MITP/20-039}
\preprint{NCTS-TH/2010}

\begin{abstract}
A non-trivial $\mathcal{S}$-matrix generally implies a production of entanglement:  starting with an incoming pure state the scattering generally returns an outgoing state with non-vanishing entanglement entropy. It is then interesting to ask if there exists a non-trivial $\mathcal{S}$-matrix that generates  no entanglement. In this letter, we argue that the answer is the scattering of classical black holes. We study the spin-entanglement in the scattering of arbitrary spinning particles. Augmented with Thomas-Wigner rotation factors, we derive the entanglement entropy from the gravitational induced $2\rightarrow 2$ amplitude. In the Eikonal limit, we find that the relative entanglement entropy, defined here as the \textit{difference} between the entanglement entropy of the \textit{in} and \textit{out}-states, is nearly zero for minimal coupling irrespective of the \textit{in}-state, and increases significantly for any non-vanishing spin multipole moments. This suggests that minimal couplings of spinning particles, whose classical limit corresponds to Kerr black hole, has the unique feature of generating near zero entanglement.   
\end{abstract}

\maketitle

\section{Introduction}

One of the fascinating realizations in the interplay of gravitational scattering amplitudes and dynamics of compact binary systems, is the equivalence of  minimally coupled spinning particle and  rotating black holes. In the analysis of three-point amplitudes of particles with general spin, a unique amplitude for massive spin-$s$ particle emitting a massless graviton,  was defined kinematically in~\cite{Arkani-Hamed:2017jhn} and termed \textit{minimal coupling}.  The term reflects its matching to minimal derivative coupling when taking the high energy limit for $s\leq 2$. Since massless particles have spins bounded by 2 in flat space, the role of minimal coupling with $s>2$ was initially not clear. 

Through a series of subsequent analysis~\cite{Guevara:2017csg, Chung:2018kqs, Guevara:2018wpp, Arkani-Hamed:2019ymq,Aoude:2020onz}, it was understood that the spin multipoles generate by minimal coupling are exactly that of a spinning black hole, i.e. the spin moments in the effective stress-energy tensor of the linearized Kerr solution. This was verified by reproducing the Wilson coefficients of one-particle effective theory (EFT)~\cite{Goldberger:2004jt, Porto:2008tb} for Kerr black hole~\cite{Chung:2018kqs}, and the classical scattering angle at leading order in the Newton constant $G$ to all orders in spin~\cite{Guevara:2018wpp}.

While the equivalence can be established through various direct matching, the principle that underlies such correspondence remains unclear. In this letter, we seek to answer this by studying the spin entanglement entropy. We will use the action of the $2\rightarrow 2$ $\mathcal{S}$-matrix in the Eikonal limit on two particle spin-states. By measuring the relative entanglement entropy for the final state, defined as
\eq\label{Relative}
\Delta S\equiv - {\rm tr} \left[\rho^{\rm out} \log \rho^{\rm out} \right] {+} {\rm tr} \left[\rho^{\rm in} \log \rho^{\rm in} \right] \,,
\eqe
where $\rho^{\rm in,out}$ is the reduced density matrix for the in and out-state, remarkably we find that $\Delta S \approx 0$ when the $\mathcal{S}$-matrix is given associated with minimal coupling, or equivalently, when the EFT Wilson coefficients are set to the black hole value, unity. Any deviation from unity significantly increases the relative entropy. 

\section{Entanglement via S-matrix}
\label{sec:entSmatrix}\vspace*{-1mm}
The study of entanglement in scattering events has a long history, which, for recent developments we refer to~\cite{Cervera-Lierta:2017tdt, Beane:2018oxh, Afik:2020onf, Bose:2020shm}. Denote the two particle Hilbert space by $\mathcal{H}=\mathcal{H}_a \otimes \mathcal{H}_b$,  for each subsystem we can further divide into spin and momentum degrees of freedom, e.g. $\mathcal{H}_{\rm a}=\mathcal{H}_{s_a} \otimes \mathcal{H}_{p_a}$. In computing the entanglement from scattering, there are two sources of difficulty. First, the trace over momentum states lead to divergences due to the infinite space-time volume, and introducing cut-off leads to regulator dependent results, see e.g.~\cite{Peschanski:2016hgk, Peschanski:2019yah}. Second, under Lorentz rotations, the spin undergoes Thomas-Wigner rotation and one does not have a Lorentz invariant definition of the reduced density matrix~\cite{PhysRevLett.88.230402, Lindner:2003ve}  (see \cite{He:2007du} for further discussions).
 
On the other hand, the same difficulty also appears in the extraction of conservative Hamiltonian of binary systems from relativistic scattering amplitudes. In particular, in a $2\rightarrow 2$ scattering process, the spin (little group) space of the incoming particles are invariably distinct from the outgoing space, as their momentum are distinct. However, by augmenting the $\mathcal{S}$-matrix with Thomas-Wigner rotation factors,  the final state can be mapped back into the spin Hilbert space of the incoming state.  Indeed such \emph{Hilbert space matching} procedure was used heavily in the computation of the spin-dependent part of the conservative Hamiltonian~\cite{Chung:2019duq, Chung:2020rrz, Bern:2020buy}.

We thus consider elastic scattering in the spin Hilbert space  $\mathcal{H}=\mathcal{H}_{s_a} \otimes \mathcal{H}_{s_b}$. With a given in-state,  we can obtain the out-state via the amplitude as:
\eq\label{e:out}
|{\rm out} \rangle=(U_a\otimes U_b )\,M\,|{\rm in}\rangle\,,
\eqe
where $U_{a,b}$ are the Hilbert space matching factors which will be discuss in the next section~\cite{Fan}. The total density matrix of the out-state is then simply  $\rho^{{\rm out}}_{a,b} = |{\rm out}\ra \la {\rm out}|$  and the reduced density matrix is given by  $\rho_{\rm a} = {\rm tr}_b \rho_{\rm a,b}$. Equipped with $\rho_{\rm a}$ we can consider a variety of entanglement quantifiers. A canonical choice is the entanglement entropy, i.e. the Von Neumann entropy of the reduced density matrix $S_{\rm VN} = - {\rm tr}_a \left[\rho_a \log \rho_a \right]$.  Note that here, $S_{\rm VN}$ in principle depends on the in-state. For a quantifier that is independent of the in-state, we can consider the entanglement power \cite{nielsen_chuang_2010} given by
\begin{align}
\label{e:EP}
\mathcal{E}_a= 1- \int \frac{d\Omega_a}{4\pi}\frac{d\Omega_b}{4\pi} {\rm tr}_a \rho_a^2.
\end{align}
where $\Omega$ represents the spin-$s$ phase space. 

In the following, we will consider the elastic $\mathcal{S}$-matrix acting on $|{\rm in}\rangle=|s_a\rangle\otimes|s_b\rangle$, i.e. the in-state is set up as a pure state. Thus by computing the entanglement entropy of the out-state, we obtain the entanglement enhancement of the scattering process.

\section{The Eikonal Amplitude in Spin Space}
In this section, we compute the leading order amplitude $a b \rightarrow a' b'$ for general massive spinning particles in the Eikonal limit. Working in the center of mass frame where $p_a = (E_a, 0, 0, \vec{p})$, $p_b = (E_b, 0, 0, -\vec{p})$, and the momentum transfer $q = p_a - p_a^{\prime} = (0, \,\vec{q})$, the Eikonal limit correspond to $q^2  \rightarrow 0$. After Fourier transform to the impact parameter space, we obtain the Eikonal phase whose exponentiation yields to the $\mathcal{S}$-matrix in the Eikonal limit.
\subsection{Spin-$s$ Amplitudes and Hilbert Space Matching}
\label{s:hilbertM}
\begin{figure}
\begin{center}
\includegraphics[scale=0.5]{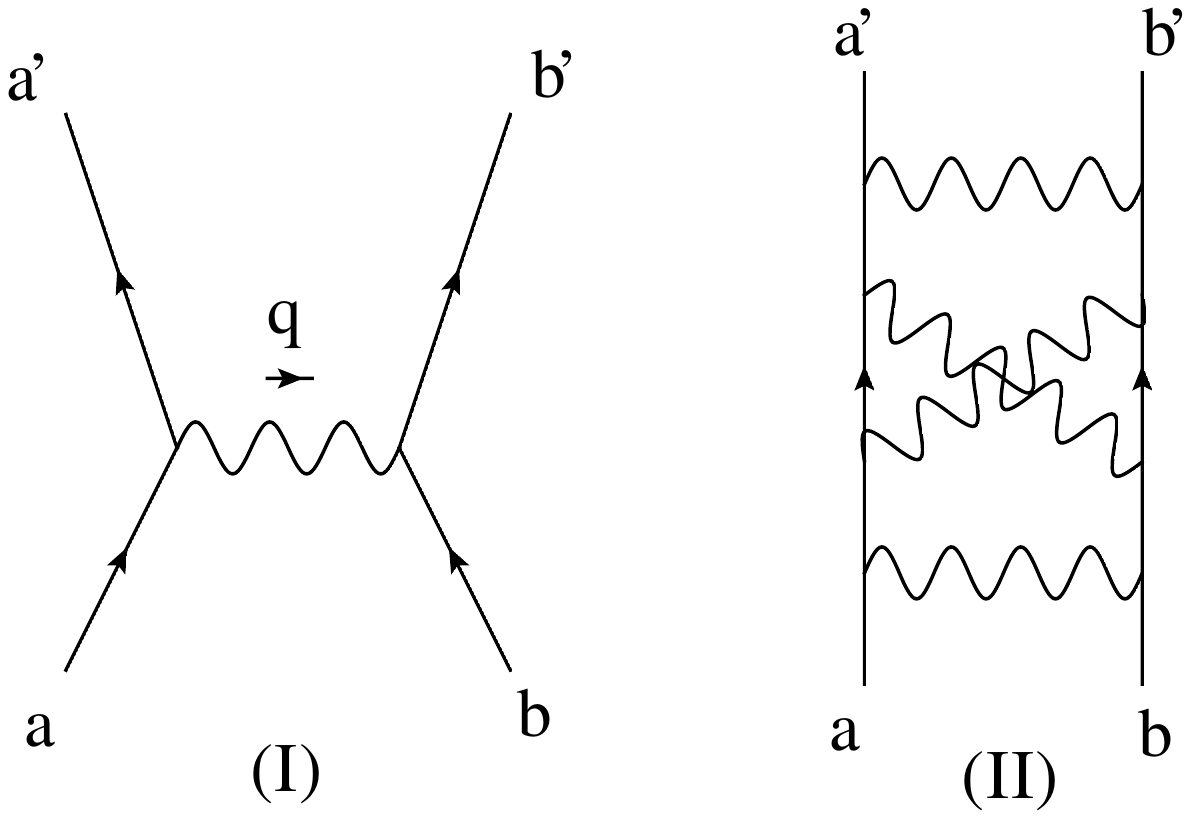}
\caption{We consider the $2\rightarrow 2$ scattering of two spinning objects exchanging gravitons. (I) Process in the leading order of the Newton constant $G$.  (II) Eikonal approximation, which re-sums the ladder diagrams.}
\label{Sfig}
\end{center}
\end{figure}

We begin with the scattering of spinning particles induced by gravitational interactions. At leading order in the Newton constant $G$, the four point amplitude for the $a b\rightarrow a' b'$ illustrated in fig. \ref{Sfig}, can be written as~\cite{Chung:2020rrz}:
\begin{align}\label{eq:4pt}
&M_{\rm{tree}}(q^2)
= {-}8\pi G \frac{m_a^2 m_b^2}{q^2} \times  \\
& \times \sum_{\eta = \pm 1} e^{2\eta\Theta}  [\varepsilon_{a^{\prime}}^* W_a(\eta \tau_a) \varepsilon_{a}] [\varepsilon_{b^{\prime}}^{*}W_2(\eta \tau_b)\varepsilon_{b}] + \mathcal{O}(q^0)\, ,\nonumber
\end{align}
where $q^\mu$ is the the transfer momenta, $\varepsilon_i$ is the polarization tensor of the spinning particle, $\tau_{a,b} = \frac{q\cdot S}{m_{a,b}}$ and the exponential parameters are defined as $\cosh\Theta \equiv \frac{p_a \cdot p_b}{m_a m_b}$  and $\eta = \pm 1$ labelling the exchanged graviton's helicity. The function $W(\eta \tau)$ is defined as:
\begin{equation}
\label{e:Wi}
W_{a,b}(\eta \tau_{a,b})=\left[ \sum_{n=0}^{2s_{a,b}} \frac{C_n}{n!} \left(\eta \frac{q\cdot S}{m_{a,b}} \right)^n\right]\,,
\end{equation}
where $S$ is the Pauli-Lubanski spin-vector and $C_{a,n},\,C_{b,n}$ parametrizes the possible distinct couplings for particle $a,b$. These are the $2s$ multi-pole moments carried by a spin-$s$ particle, and can be directly matched to the Wilson coefficients of the one-particle effective action (see \cite{Levi:2015msa} for the all order in spin action). For rotating black holes $C_{a,n}=C_{b,n}=1$, and the classical spin is recovered in the limit $s\rightarrow \infty$, $\hbar\rightarrow 0$ while keeping the classical spin $S\equiv s \hbar$ fixed (see \cite{Maybee:2019jus} for a more detailed discussion).

As shown in ref.~\cite{Chung:2019duq}, we can transform the spin-vector $S$ in an operator acting in the little group space through the insertion of a complete set of polarization tensors associated to the incoming particles:
\begin{align}
\mathbb{S}_{a,b} \equiv \varepsilon_{a,b, \{I_s\}}^* S^{\mu} \varepsilon_{a,b}^{\{J_s\}}\, ,
\end{align} 
where $\{I_s\},\,\{J_s\}$ are the $SU(2)$ indices of particle $a,\,b$. In components, we have that
\begin{align}\label{eq:S component}
\mathbb{S}^{\mu}_{a,b} = \left(
\frac{\vec{p}_{a,b} \cdot \vec{\Sigma}}{m_{a,b}},~ \vec{\Sigma} + \frac{\vec{p}_{a,b}\cdot \vec{\Sigma}}{m_{a,b}(m_{a,b}+E_{a,b})}\vec{p}_{a,b}\right),
\end{align}
where $\vec{\Sigma}$ is the spin-$s$ rest frame spin operator satisfying the commutation relation $[\Sigma_i, \Sigma_j] = i \epsilon_{ijk}\Sigma_k$. Then the operator $\tau$ in the little group space is given by
\begin{align}
\mathbb{T}_{a,b} \equiv \varepsilon_{a,b, \{I_s\}}^* \frac{q\cdot S}{m_{a,b}} \varepsilon_{a,b}^{\{J_s\}} \equiv \frac{q\cdot \mathbb{S}_{a,b}}{m_{a,b}}.
\end{align}
Writing eq.~\eqref{eq:4pt} in term of $\mathbb{T}$ leads to an amplitude that corresponds to an operator acting on states in distinct little group space,  as the momenta of $a,b$ are distinct from $a',b'$.  This can be rectified by the so called \textit{Hilbert space matching} procedure which utilize the Lorentz transformation that relates the momenta of the in-states to the out-states, to convert the out-states Hilbert space back to the in-states~\cite{Chung:2019duq, Chung:2020rrz}. The result is the additional Thomas-Wigner rotation factors for each of the two particles. For example, for particle $a$ this factor, in leading order in $q^2$, is written as
\begin{equation}
U_a = \exp\left[-i \frac{m_a m_b \,\mathbb{E}_a }{(m_a + E)E}  \right],
\end{equation}
where $\mathbb{E}_a \equiv \epsilon(q, u_a, u_b, a_a) = \epsilon_{\mu\nu\rho\sigma}q^{\mu}u_a^{\nu}u_b^{\rho}a_a^{\sigma}$, $a_a = \mathbb{S}_a/m_a$, $u_{a,b} = p_{a,b}/m_{a,b}$ and $E = E_a + E_b$. In summary, the amplitude after the Hilbert space matching, denoted by  $\overline{M}$,  is given by
\begin{align}\label{eq:q Amplitude}
\overline{M}_{\rm tree}&(q^2) =
 -8\pi G \frac{m_a^2 m_b^2}{q^2}\times \\ 
 &\times \sum_{\eta = \pm 1} e^{2\eta\Theta} W_a(\eta \mathbb{T}_a) W_b(\eta \mathbb{T}_b) U_a U_b + \mathcal{O}(q^0)\,. \nonumber
\end{align}

Expanding eq.~\eqref{eq:q Amplitude} up to order $\mathcal{O}(\mathbb{S}^{2s_i})$ gives
\begin{align}\label{eq:A basis}
&\overline{M}_{{\rm tree}}(q^2) = -\frac{16\pi Gm_a^2 m_b^2}{q^2} \\
&\times \left\lbrace \sum_{m=0}^{\floor{s_a}} \sum_{n=0}^{\floor{s_b}} A_{2m,2n} \left( \mathbb{T}_{a}^{2m} \otimes \mathbb{T}_{b}^{2n}\right) \right. \nonumber \\
\quad
&+ \frac{m_a^2 m_b}{E} 
\sum_{m=0}^{\ceil{s_a}-1} 
\sum_{n=0}^{\floor{s_b}} A_{2m+1, 2n}  \left(\text{Sym} \left[\mathbb{E}_a \mathbb{T}_{a}^{2m} \right]\otimes \mathbb{T}_{b}^{2n} \right)\nonumber \\
\quad
&+ \frac{m_a m_b^2}{E}
\sum_{m=0}^{\floor{s_a}}\sum_{n=0}^{\ceil{s_b}-1} A_{2m, 2n+1} \left(\mathbb{T}_{a}^{2m} \otimes \text{Sym} \left[\mathbb{E}_b \mathbb{T}_{b}^{2n} \right]  \right) \nonumber \\
&+
\left.\sum_{m=0}^{\ceil{s_a}-1} \sum_{n=0}^{\ceil{s_b}-1}
A_{2m+1, 2n+1} \left(\mathbb{T}_{a}^{2m+1}  \otimes \mathbb{T}_{b}^{2n+1}\right)
\right\rbrace \nonumber \, ,
\end{align}
where  we used the shorthand notation $\mathbb{T}_{a,b} \equiv \left(q\cdot a_{a,b}\right)$ and 
\begin{align}
&\text{Sym} \left[\mathbb{E}_{i} \mathbb{T}_{i}^{2n} \right] \equiv  \\ 
&\frac{1}{2n+1} \left[\mathbb{E}_i\mathbb{T}_{i}^{2n} + \mathbb{T}_{i}\mathbb{E}_i  \mathbb{T}_{i}^{2n-1} + \cdots + \mathbb{T}_{i}^{2n} \mathbb{E}_i \right] \nonumber \,,
\end{align}
for $i=a,b$. The explicit form of the coefficients $A_{m,n}$ in eq.~\eqref{eq:A basis}, up to $m,n=2$, is given by
\begingroup
\allowdisplaybreaks
\begin{align}
A_{0,0} &= c_{2\Theta}\,, \quad A_{1,0} = \frac{i(2E r_a c_{\Theta} - m_b c_{2\Theta})}{m_a^2 m_b r_a} \,, \\
A_{1,1} &= \frac{c_{2 \Theta } s_{\Theta }^2 m_a m_b}{E^2 r_a r_b}+c_{2 \Theta }-\frac{2 m_b c_{\Theta } s_{\Theta }^2}{E  r_a}-\frac{2 m_a c_{\Theta }s_{\Theta }^2}{E r_b}\,, \nonumber \\
A_{2,0} &=\frac{C_{a,2} c_{2 \Theta }}{2}+\frac{m_b^2 c_{2 \Theta } s_{\Theta }^2}{2 E^2  r_a^2}-\frac{2 m_b c_{\Theta } s_{\Theta }^2}{E r_a}\,,\nonumber  \\
\begin{split}
A_{2,1} &= 
i \left( \frac{E C_{a,2} c_{\Theta }}{m_a m_b^2}
-\frac{C_{a,2} c_{2 \Theta }}{2  m_b^2 r_b}
+\frac{c_{2 \Theta }}{4E^2  r_a^2r_b}
\right.
\\
&
\left.
-\frac{c_{4 \Theta }}{8 E^2  r_a^2 r_b}
-\frac{c_{\Theta }}{2 E r_a m_b r_b}
+\frac{c_{3 \Theta }}{2 E  r_a m_br_b}
\right.
\\
&
\left.
-\frac{c_{2 \Theta }}{m_a r_a m_b}
-\frac{1}{8 E^2  r_a^2 r_b}
-\frac{c_{\Theta }}{4 E m_a r_a^2}
+\frac{c_{3 \Theta }}{4 E m_a r_a^2}
\right)\,,
\end{split}
\nonumber\\
\begin{split}
A_{2,2} &= 
\frac{C_{a,2} C_{b,2} c_{2 \Theta }}{4}
-\frac{C_{a,2} c_{\Theta } s_{\Theta }^2 m_a}{E r_b}
-\frac{C_{b,2} c_{\Theta }s_{\Theta }^2 m_b}{E r_a }\\
&
+\frac{c_{2 \Theta } s_{\Theta }^4m_a^2 m_b^2}{4 E^4 r_a^2 r_b^2}
-\frac{c_{\Theta } s_{\Theta }^4m_a m_b^2}{E^3 r_a^2r_b}
-\frac{c_{\Theta } s_{\Theta }^4m_a^2 m_b}{E^3 r_a  r_b^2}
\\
&
+\frac{c_{2 \Theta } s_{\Theta }^2m_a m_b}{E^2  r_a  r_b}
+\frac{C_{b,2} c_{2 \Theta }s_{\Theta }^2 m_b^2}{4E^2  r_a^2}
+\frac{C_{a,2} c_{2 \Theta } s_{\Theta }^2 m_a^2 }{4 E^2 r_b^2}\,,
 \nonumber
\end{split}
\end{align}
\endgroup
where $C_{a,2}$ and $C_{b,2}$ are the Wilson coefficients for each particles, $(c_{\Theta}, s_{\Theta} )\equiv (\cosh \Theta, \sinh \Theta)$ and $r_{a,b} \equiv 1+ E_{a,b}/m_{a,b}$. We can see that the Wilson coefficients $C_{a,n}$ and $C_{b,n}$ starts to appear at $A_{2,0}$, which means that we need to go to at least to spin-1 to compare the difference between black holes and other objects.

\subsection{Eikonal Phase}
The Eikonal phase, at order $\mathcal{O}(G)$, is given simply by the Fourier transform of the tree-level amplitude in eq.~\eqref{eq:q Amplitude} to the impact parameter space:
\eq\label{eq:Eikonal phase}
\chi(b)= \frac{1}{4|\vec{p}|E}\int \frac{d^2\vec{q}}{(2\pi)^2}\;\; e^{i \vec{q}\cdot \vec{b}}\overline{M}_{{\rm tree}}(q^2)\,.
\eqe
Since $q^2 \rightarrow 0$ in the Eikonal limit, we have $\vec{q}\cdot\vec{p} = q^2/2 \rightarrow 0$. This orthogonality between $\vec{q}$ and $\vec{p}$ defines the impact parameter space, which is the plane perpendicular to the incoming momentum, i.e. $\vec{b} = (b_x, b_y, 0)$. 
Note that, in this limit, we can simply replace all $\vec{\mathbb{S}}$  in eq.~\eqref{eq:A basis} by $\vec{\Sigma}$, which is the rest frame spin operator. The $\mathcal{S}$-matrix in the Eikonal approximation is then the exponential of the phase: 
\begin{equation}
\mathcal{S}_{{\rm Eikonal}}= e^{i\chi(b)}\,.
\end{equation}
This allow us to write the out-state in the Eikonal approximation, replacing the matrix element  of $U_a U_b\,M$ by the ones of $\mathcal{S}_{\rm Eikonal}$ in eq.~\eqref{e:out}:
\eq
\label{e:outEik}
|{\rm out}\rangle =\mathcal{S}_{\rm Eikonal}| {\rm in}\rangle\,.
\eqe
\section{The entanglement entropy of binary systems}

\begin{figure}
  \centering
  \begin{tabular}{@{}c@{}}
    \includegraphics[width=8cm]{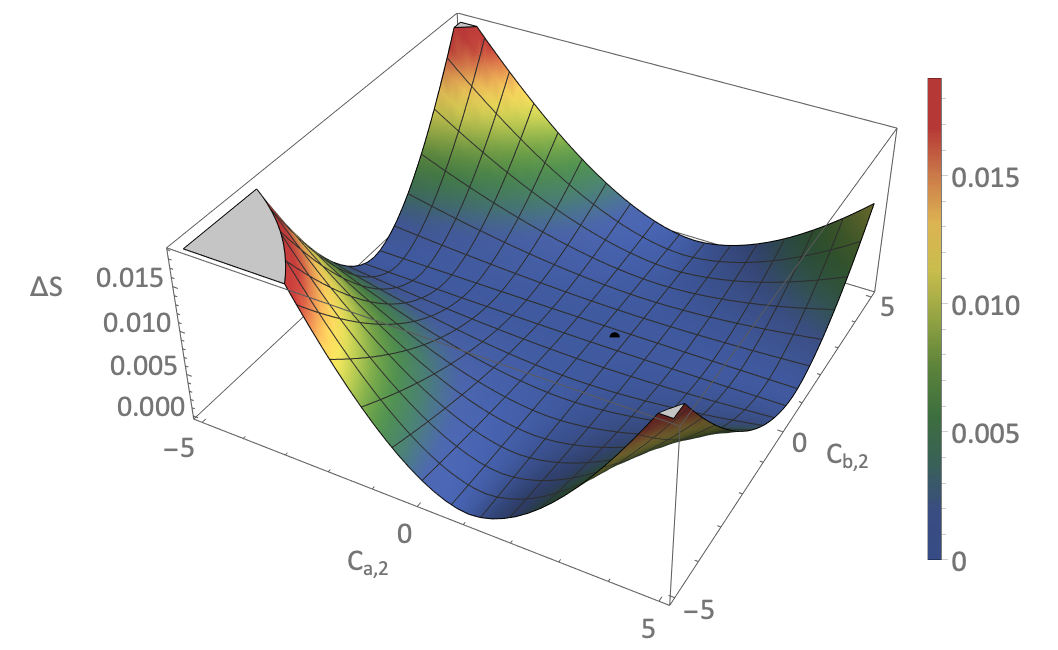} \\[\abovecaptionskip]
    \small (I) Relative Von Neumann entropy.

  \end{tabular}

  \vspace{\floatsep}

  \begin{tabular}{@{}c@{}}
    \includegraphics[width=8cm]{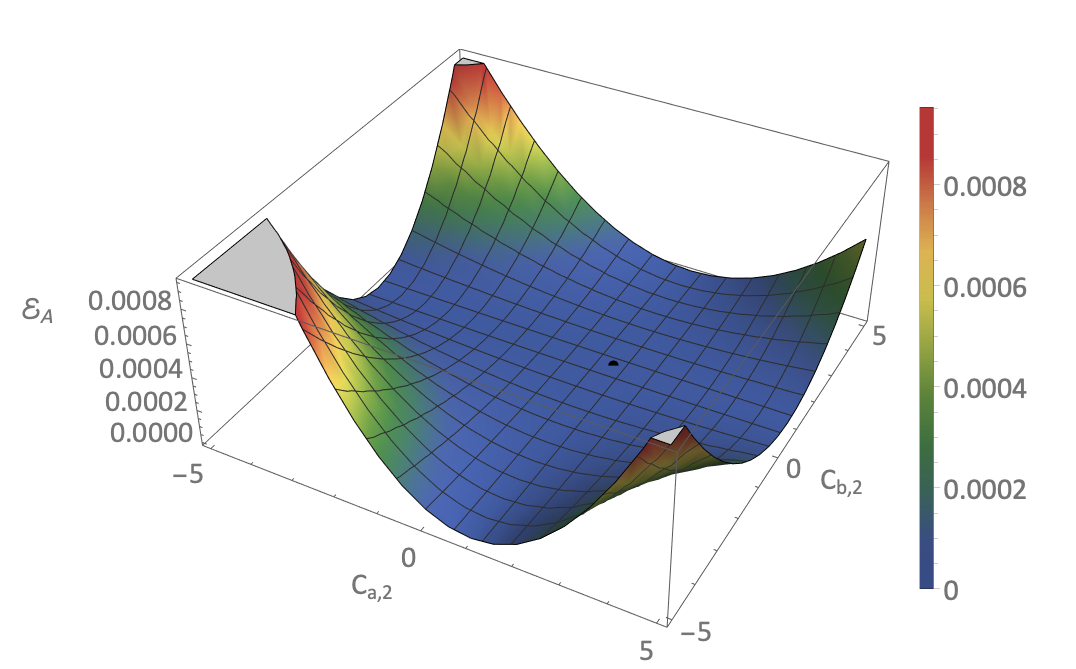} \\[\abovecaptionskip]
    \small (II) Entanglement power.
  \end{tabular}

\caption{ (I) Relative entanglement entropy $\Delta S$ and (II) the entanglement power $\mathcal{E}_a$ for massive spin-1 particles. The initial state is set to $\lvert{\rm in}\ra=\lvert\upuparrows\ra$ and the kinematic parameters are given by $|\vec{p}_a| = |\vec{p}_b| = |\vec{p}|$, $m_a = m_b = m$, $\vec{b} = (b, 0,0)$, $Gm^2 = 10^{-4}$, $|\vec{p}|b = 1000$, $|\vec{p}|/m = 100$. The minimum, represented by the black point, corresponds to the Wilson coefficient value $(C_{a,2}, C_{b,2})=(1,1)$, $\Delta S \approx 1.54\times10^{-9}$ and $\mathcal{E}_a \approx 1.10\times10^{-10}$. }\label{fig:spin1}
\end{figure}

We now have all the ingredients necessary to compute the entanglement entropy and the entanglement power for the out-state in the Eikonal approximation. We first compute the entanglement entropy for spin-$1$ particles, which corresponds to keep spin operators up to second power for each particle in the Eikonal phase. Starting with a pure state $\lvert{\rm in}\ra=\lvert\upuparrows\ra$, the entanglement entropy for the resulting out-state yields directly to the relative entropy $\Delta S$ in eq.~(\ref{Relative}). The result is plotted in fig.~\ref{fig:spin1} against the Wilson coefficients pair $(C_{a,2}, C_{b,2})$. Remarkably,  the minimum is exactly at the Kerr black hole value $C_{a,2}, =C_{b,2}=1$ and deviating from this point raises the entropy of the system. This is unchanged for different choice of in-states, which is illustrated by the computation of entanglement power given by eq.~\eqref{e:EP} and shown in fig.~\ref{fig:spin1}.  We've also obtained similar result with mixed instates. 

In order to show that this is indeed a robust result, we also consider higher spins. Using the same set up we calculate the relative Von Neumann entropy for spin-3 massive particles, which has a total of $5+5=10$ Wilson coefficients. In our extensive scan, we find that the black hole value, $C_{a,i}=C_{b,i}=1$ for $i=2,\cdots,6$, is the unique point that gives the minimum value. As an illustrative example, we set all Wilson coefficients to one except the pair $(C_{a,2}, C_{b,2})$ and plot $\Delta S$ with respect to $(C_{a,2}, C_{b,2})$ in fig.~\ref{fig:spin3}. The results show the minimum at $(1,1)$, while the two orthogonal valleys represent keeping only one of the coefficient at one. In fig.~\ref{fig:spin3ca2ca3}, we plot $C_{a,2}=C_{b,2}=C_2$ and $C_{a,3}=C_{b,3}=C_3$, while keeping all remaining coefficients one. Once again the corresponding black hole point gives near zero entanglement. 

\begin{figure}
 \includegraphics[width=8cm]{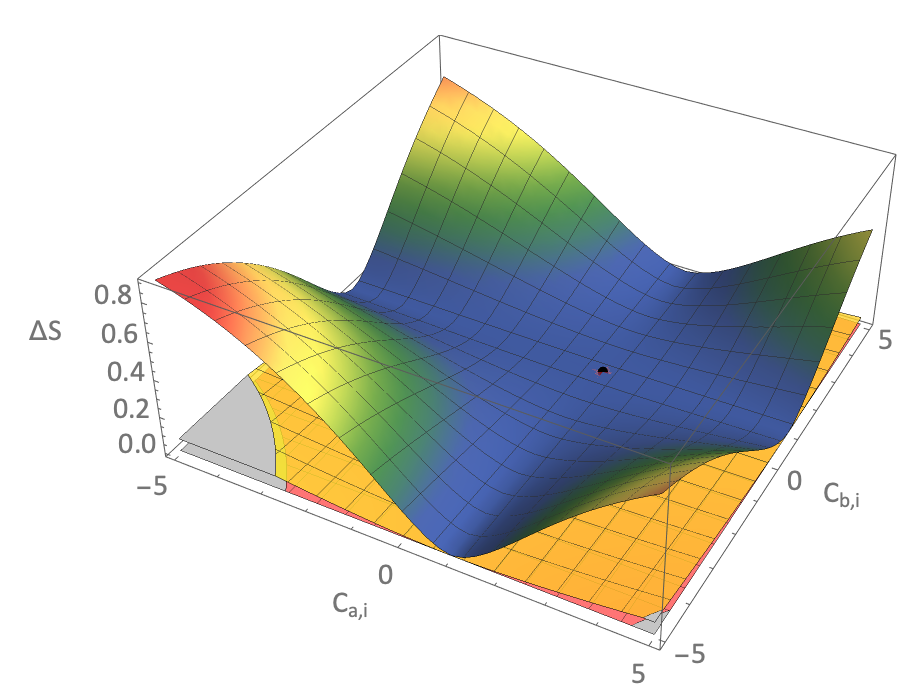}
 \centering
 \caption{Relative entanglement entropy for massive spin-2 particles. The initial state is set to $\lvert{\rm in}\ra=\lvert\upuparrows\ra$ and the kinematic parameters are given by $|\vec{p}_a| = |\vec{p}_b| = |\vec{p}|$, $m_a = m_b = m$, $\vec{b} = (b, 0,0)$, $Gm^2 = 10^{-4}$, $|\vec{p}|b = 1000$, $|\vec{p}|/m = 100$. The planes corresponds respectively to  $(C_{a,2},C_{b,2})$, $(C_{a,3},C_{b,3})$ and $(C_{a,4},C_{b,4})$, while all others Wilson coefficients are set to one.  In any of the cases, the minimum, represented by the black point, corresponds to the Wilson coefficients set to one and  $\Delta S \approx 5.84\times10^{-9}$.}
 \label{fig:spin2}
\end{figure}

\begin{figure}
 \includegraphics[width=8cm]{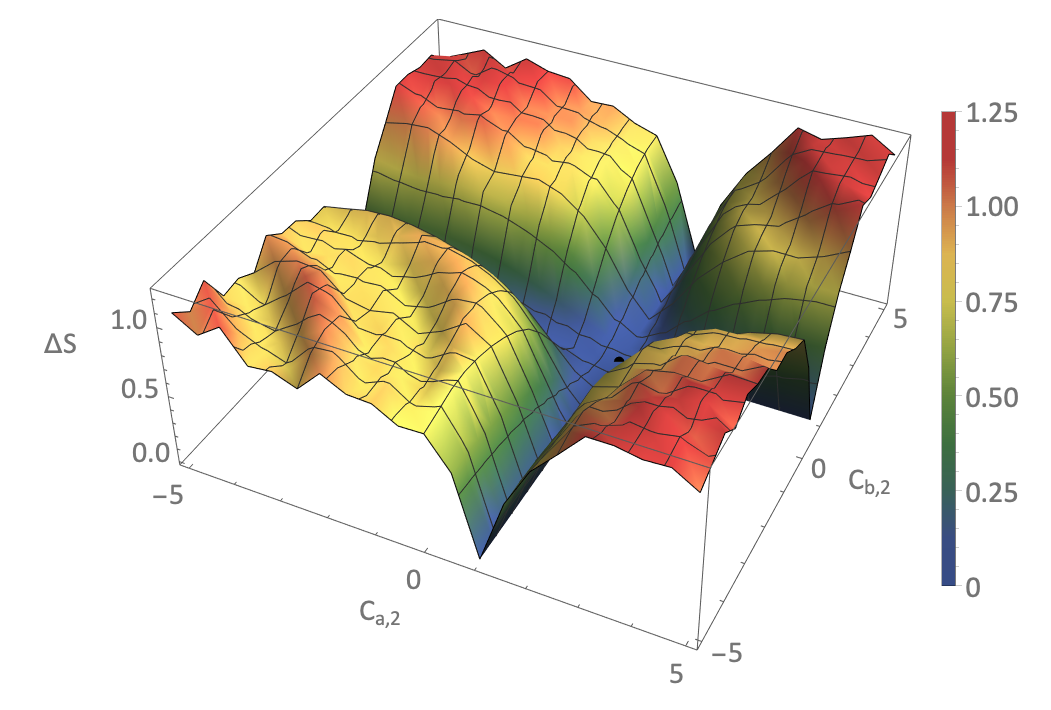}
 \centering
 \caption{Relative entanglement entropy for massive spin-3 particles. The initial state is set to $\lvert{\rm in}\ra=\lvert\upuparrows\ra$ and the kinematic parameters are give by $|\vec{p}_a| = |\vec{p}_b| = |\vec{p}|$, $m_a = m_b = m$, $\vec{b} = (b, 0,0)$, $Gm^2 = 10^{-4}$, $|\vec{p}|b = 1000$, $|\vec{p}|/m = 100$. The Wilson coefficients  $(C_{a, i \neq 2},C_{b, j \neq 2})$ are set to one. The minimum, represented by the black point, is at  $\Delta S \approx 1.26\times10^{-8}$ and corresponds to the Wilson coefficient value $(C_{a,2},C_{b,2})=(1,1)$.}
 \label{fig:spin3}
\end{figure}

\begin{figure}
 \includegraphics[width=8cm]{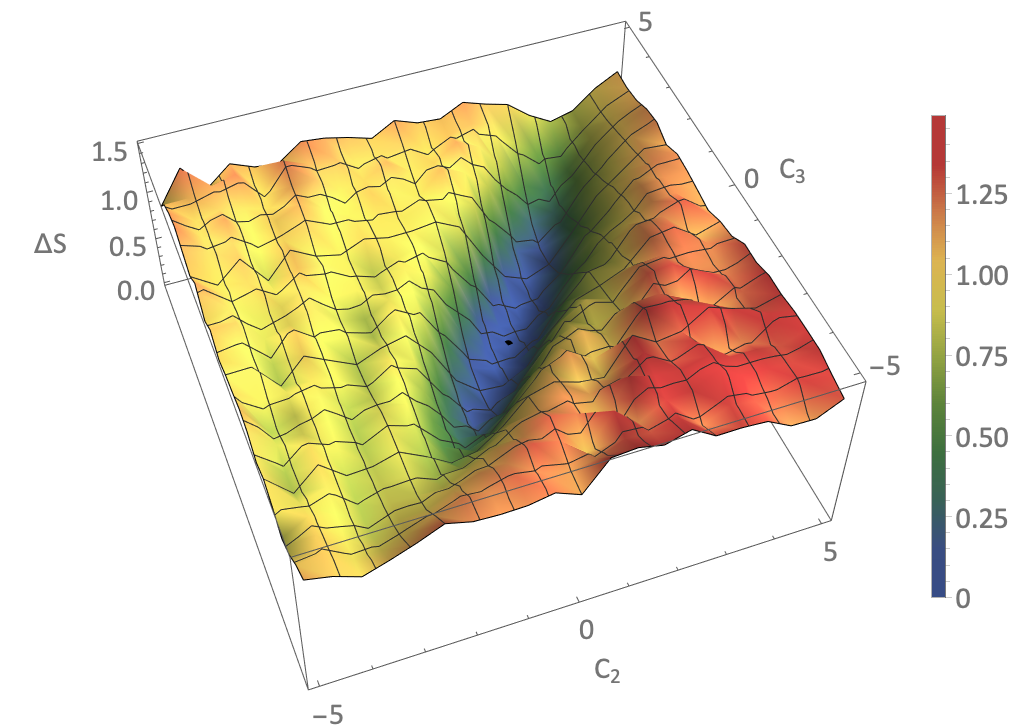}
 \centering
 \caption{Relative entanglement entropy for massive spin-3 particles. The initial state is set to $\lvert{\rm in}\ra=\lvert\upuparrows\ra$ and the kinematic parameters are given by $|\vec{p}_a| = |\vec{p}_b| = |\vec{p}|$, $m_a = m_b = m$, $\vec{b} = (b, 0,0)$, $Gm^2 = 10^{-4}$, $|\vec{p}|b = 1000$, $|\vec{p}|/m = 100$, $C_{a,2}=C_{b,2}=C_2$, $C_{a,3}=C_{b,3}=C_3$. All others Wilson coefficients  are set to one. The minimum is at $\Delta S \approx 1.26\times10^{-8}$ and corresponds to the Wilson coefficient value $(C_{2},C_{3})=(1,1)$ . }
 \label{fig:spin3ca2ca3}
 \end{figure}

While the deformation of each Wilson coefficient away from the unity  raises the entanglement entropy, comparatively, the effect of $C_{2}$ is dominant. This is illustrated in fig.~\ref{fig:spin2} that compares $\Delta S$ for deforming the three different pairs of Wilson coefficients in the spin-2 system.  We can observe that deforming $(C_{a, 2},C_{b,2})$ has the dominant effect in generating entanglement.  

\begin{figure}
 \includegraphics[width=8cm]{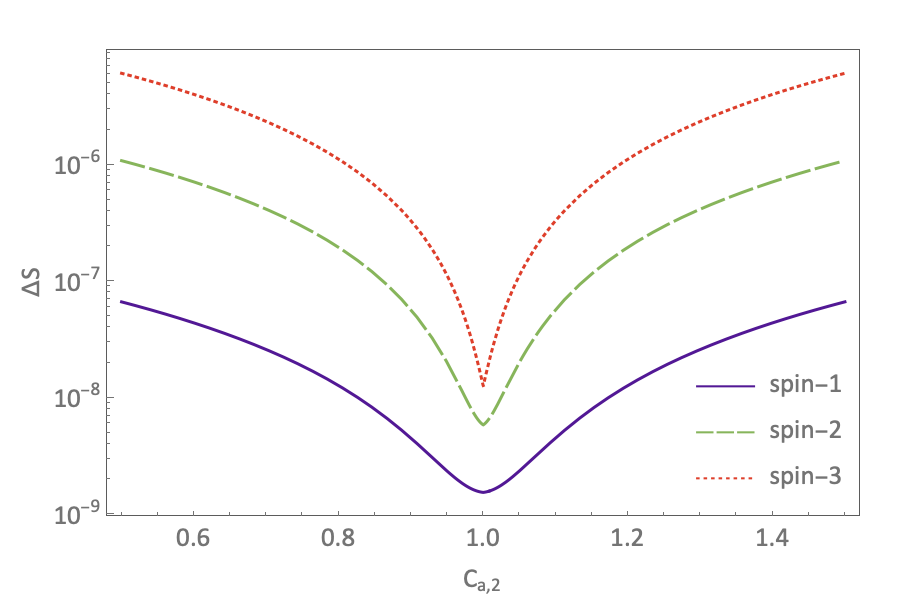}
 \centering
 \caption{Comparison between the relative entanglement entropy for spin-1, spin-2 and spin-3. All Wilson coefficients are set to one except $C_{a,2}$. }
 \label{fig:1d}
\end{figure}

Finally, we expect that including higher spins do not change the main result. The minimum of the relative entropy is always at the Kerr black hole Wilson coefficient point. Moving away from this point quickly increase the entanglement entropy. A comparison between the spin-1, spin-2 and spin-3 cases keeping all Wilson coefficients one except $C_{a,2}$ can be seen in fig.~\ref{fig:1d}.\\

\section{Conclusions and outlook}
In this letter, we consider the entanglement entropy generated by gravitationally coupled binary systems. By considering the Hilbert space of spin states, we demonstrate that minimal coupling for massive arbitrary spin particle have the unique feature of generating nearly zero entanglement in the scattering process. Given the correspondence between minimal coupling and rotating black holes, the result suggests that such feature can also be attributed to the entanglement properties of spinning black holes.  Note that such phenomenon is reminiscent of what was found in strong interactions, where entanglement suppression is associated with symmetry enhancement~\cite{Beane:2018oxh}.

While the relative entropy is near zero, it is not zero, which may be an artifact of confining ourselves to leading order in Eikonal approximation. This makes clear investigation at NLO is desirable.  As mentioned in the introduction, there is a general correspondence between minimal coupling and black-hole like solutions in four-dimensions. This includes Reissner Nordstrom, Kerr Newman~\cite{Moynihan:2019bor, Chung:2019yfs}, Taub-NUT~\cite{Huang:2019cja} and Kerr-Taub NUT~\cite{NewPaper}. Furthermore, gravitationally induced spin-multipoles has also been studied recently in the context of fuzzball microstates~\cite{Bena:2020see, Bena:2020uup, Bianchi:2020bxa}. For Kerr Newman there are additional electromagnetic spin multipoles, while for Kerr-Taub NUT and fuzzballs, the minimal couplings are dressed with additional complex phase factors. It will be fascinating to explore their features through the prism of spin entanglement.  Finally, it will also be interesting to understand quantum corrections, in particular whether or not they generate anomalous gravitational multipole moments.

\section*{Acknowledgements}
We would especially like to thank Jung-Wook Kim, for discussions on the computation of Eikonal phase for spin effects. Also Bo-Ting Chen, Tzu-Chen Huang, Jun-Yu Liu and Andreas Helset for enlightening discussions.  C.S.M. thanks Hugo Marrochio for the encouragement in the early stages of the project. 
The work of C.S.M.\ and R.A.\ is supported by the Alexander von Humboldt Foundation, in the framework of the Sofja Kovalevskaja Award 2016, endowed by the German Federal Ministry of Education and Research and also supported by  the  Cluster  of  Excellence  ``Precision  Physics,  Fundamental Interactions, and Structure of Matter'' (PRISMA$^+$ EXC 2118/1) funded by the German Research Foundation (DFG) within the German Excellence Strategy (Project ID 39083149). Mz C, Yt H, and Mk T is supported by MoST Grant No. 106-2628-M-002-012-MY3. Yt H is also supported by Golden
Jade fellowship.

\vspace{1.8cm}
\bibliographystyle{apsrev4-1_title}
\bibliography{refs.bib}

\clearpage
\onecolumngrid
\begin{center}
\large\textbf{Supplemental Material}
\end{center}

\section{Little group spin operators}

Since the spin vector appearing in the eikonal phase is the little group spin operator $\mathbb{S}^{\mu}$, we work out its components here by plugging in explicit parametrizations of all the variables. We parameterize the momentum $p$ as
\begin{equation}
p = (E_a, |\vec{p}|\sin\theta \cos\phi, |\vec{p}|\sin\theta \sin\phi, |\vec{p}|\cos\theta)
\end{equation}
where the corresponding spinors are given by\cite{Kim:2020ust}
\begin{equation}\label{eq:n direction spinor}
\begin{split}
\lambda_{\alpha}^{\phantom{I} I} &= (\lambda_{\alpha}^{\phantom{I} 1}, \lambda_{\alpha}^{\phantom{I} 2})\\
\lambda_{\alpha}^{\phantom{I} 1} &= 
\sqrt{E-|\vec{p}|}\ket{\hat{n},+}\braket{\hat{n},+ | \hat{z},+} + \sqrt{E+|\vec{p}|}\ket{\hat{n},-}\braket{\hat{n},- | \hat{z},+}
\\
\lambda_{\alpha}^{\phantom{I} 2} &= 
\sqrt{E-|\vec{p}|}\ket{\hat{n},+}\braket{\hat{n},+ | \hat{z},-} + \sqrt{E+|\vec{p}|}\ket{\hat{n},-}\braket{\hat{n},- | \hat{z},-}
\\
\tilde{\lambda}^{\dot{\alpha} I} &= (\tilde{\lambda}^{\dot{\alpha} 1}, \tilde{\lambda}^{\dot{\alpha} 2})\\
\tilde{\lambda}^{\dot{\alpha}1} &= 
\sqrt{E+|\vec{p}|}\ket{\hat{n},+}\braket{\hat{n},+ | \hat{z},+} + \sqrt{E-|\vec{p}|}\ket{\hat{n},-}\braket{\hat{n},- | \hat{z},+}
\\
\tilde{\lambda}^{\dot{\alpha}2} &= 
\sqrt{E+|\vec{p}|}\ket{\hat{n},+}\braket{\hat{n},+ | \hat{z},-} + \sqrt{E-|\vec{p}|}\ket{\hat{n},-}\braket{\hat{n},- | \hat{z},-}
\end{split}
\end{equation}
with the ket vectors being the eigenvectors of $\hat{n}\cdot \vec{\sigma}$:
\begin{equation}
\ket{\hat{n},+} = 
\begin{pmatrix}
\cos\left( \frac{\theta}{2}\right) \\ \sin\left( \frac{\theta}{2}\right)e^{i\phi}
\end{pmatrix}
,
\;
\ket{\hat{n},-} = 
\begin{pmatrix}
-\sin\left( \frac{\theta}{2}\right)e^{-i\phi} \\ \cos\left( \frac{\theta}{2}\right)
\end{pmatrix}
\end{equation}

\paragraph{Spin $\frac{1}{2}$}
The little group spin operator is 
\begin{equation}
(\mathbb{S}^{\mu})_{K}^{\phantom{A}I} = 
\frac{1}{2m} \bar{u}_{1,K}S^{\mu}u_1^I = -\frac{1}{2m} \AB{\lambda_K |S_{\chi}^{\mu}| \lambda^I } +  \frac{1}{2m}\SB{\tilde{\lambda}_K |S_{\bar{\chi}}^{\mu}| \tilde{\lambda}^I }
\end{equation}
with $S_{\chi}^{\mu}$ and $S_{\bar{\chi}}^{\mu}$ being the chiral and anti-chiral part of the Pauli-Lubasnki operator. Plugging in the explicit form of the spinors in eq.\eqref{eq:n direction spinor}, one would get
\begin{equation}
(\mathbb{S}^{\mu})_{K}^{\phantom{A}I} 
=
\left( \frac{\vec{p} \cdot \vec{\sigma}}{2m}, \frac{\vec{\sigma}}{2} + \frac{\vec{p} \cdot \vec{\sigma}}{2(m+E)m} \vec{p} \right)
\end{equation}
which matches with the first line of eq.(2.17) in \cite{Bern:2020buy} when we set their $\BS{S} = \frac{\vec{\sigma}}{2}$.

\paragraph{Spin 1}
For spin 1, the little group spin operator is given by
\begin{equation}
\begin{split}
\left(\mathbb{S}^{\mu}\right)_{b}^{\phantom{a}a} 
&=
 -\frac{1}{2m} (\varepsilon_{b, \beta})^* \epsilon^{\mu \nu\rho \sigma}p_{\nu}(J_{\rho\sigma})^{\beta}_{\alpha}  (\varepsilon^{a, \alpha}) \\
&=  
- \frac{i}{m}\epsilon^{\mu \nu\rho \sigma}p_{\nu} (\varepsilon_{\rho}^a)(\varepsilon_{\sigma,b}^*)
\end{split}
\end{equation}
where $\varepsilon^a_{\mu}$ is the massive spin 1 polarization vector and $a, b = -1, 0, 1$. It relates to the spinor helicity variables by:
\begin{equation}
\begin{split}
\varepsilon^{+1}_{\mu} &= \frac{\MixLeft{\lambda^1}{\sigma_\mu}{\lambda^1}}{\sqrt{2}m} \\
\varepsilon^{0}_{\mu} &= \frac{\MixLeft{\lambda^1}{\sigma_\mu}{\lambda^2} + \MixLeft{\lambda^2}{\sigma_\mu}{\lambda^1}}{2m} \\
\varepsilon^{-1}_{\mu} &= \frac{\MixLeft{\lambda^2}{\sigma_\mu}{\lambda^2}}{\sqrt{2}m} \\
\end{split}
\end{equation}
Plugging the parametrization given by eq.\eqref{eq:n direction spinor} into the above equation, we would get 
\begin{equation}
\left(\mathbb{S}^{\mu}\right)_{b}^{\phantom{a}a} =  \Big(\;\frac{\vec{p} \cdot \vec{\Sigma}}{m}, \vec{\Sigma} + \frac{\vec{p}\cdot \vec{\Sigma}}{m(m+E)}\vec{p} \;\Big)
\end{equation}
where 
\begin{equation}
\Sigma^1 = \frac{1}{\sqrt{2}}
\begin{pmatrix}
0 & 1 & 0 \\ 1 & 0 & 1 \\0 & 1 & 0 
\end{pmatrix},
\;
\Sigma^2 = \frac{1}{\sqrt{2}}
\begin{pmatrix}
0 & -i & 0 \\ i & 0 & -i \\0 & i & 0 
\end{pmatrix},
\;
\Sigma^3 = 
\begin{pmatrix}
1 & 0 & 0 \\ 0 & 0 & 0 \\0 & 0 & -1 
\end{pmatrix}
\end{equation}
We see again that our little group operator matches the first line of eq.(2.17) in \cite{Bern:2020buy} when $\BS{S} = \vec{\Sigma}$.

\section{Fourier transforming into impact parameter space}
\begin{align}
\int \frac{d^2 \vec{q}}{(2\pi)^2} e^{i\vec{q}\cdot \vec{b}} \frac{1}{|\vec{q}|^2} &= -\pi \log |\vec{b}|^2 \nonumber\\
\int \frac{d^2 \vec{q}}{(2\pi)^2} e^{i\vec{q}\cdot \vec{b}} \frac{q^i}{|\vec{q}|^2} &= i\pi \frac{2b^i}{|\vec{b}|^2} \nonumber\\
\int \frac{d^2 \vec{q}}{(2\pi)^2} e^{i\vec{q}\cdot \vec{b}} \frac{q^i q^j}{|\vec{q}|^2} &= \pi \frac{2\delta^{ij}|\vec{b}|^2 - 4 b^i b^j}{|\vec{b}|^4}\nonumber\\
\int \frac{d^2 \vec{q}}{(2\pi)^2} e^{i\vec{q}\cdot \vec{b}} \frac{q^i q^j q^k}{|\vec{q}|^2} &= -i\pi
\left[
\frac{-4(\delta^{ij} b^{k} + \delta^{ik}b^{j} + \delta^{jk}b^{l})}{|\vec{b}|^4} + \frac{16b^i b^j b^k}{|\vec{b}|^6}\right]
\\
\begin{split}
\int \frac{d^2 \vec{q}}{(2\pi)^2} e^{i\vec{q}\cdot \vec{b}} \frac{q^i q^j q^k q^l}{|\vec{q}|^2} &= -\pi
\left[
\frac{-4(\delta^{ij} \delta^{kl} + \delta^{ik}\delta^{jl} + \delta^{il}\delta^{jk})}{|\vec{b}|^4}
\right.
\\
&\phantom{12345}
\left.
+ \frac{16(\delta^{ij}b^kb^l + \delta^{jk}b^ib^l + \delta^{ki}b^jb^l + \delta^{il}b^j b^k + \delta^{jl}b^i b^k + b^i b^j \delta^{kl})}{|\vec{b}|^6}
\right.
\nonumber\\
&\phantom{12345}
\left.
-96 \frac{b^ib^jb^kb^l}{|\vec{b}|^8}
\right]
\end{split}
\end{align}

\newpage
\section{Data and More Plots}
Here we show the plots for setting $C_{a,i} = C_{b,i} = C_i$ and scan through different combinations of $C_i$ and $C_j$ to show minimal increment of entropy in a binary black hole system is robust.

\begin{figure}[H]
\begingroup
\allowdisplaybreaks
\includegraphics[width=6cm,height=6cm]{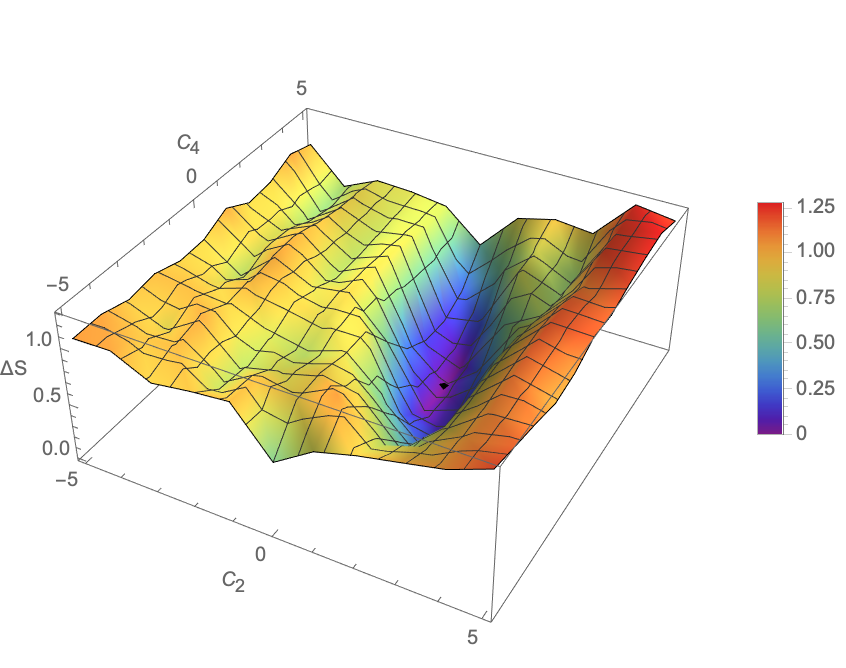}
\includegraphics[width=6cm,height=6cm]{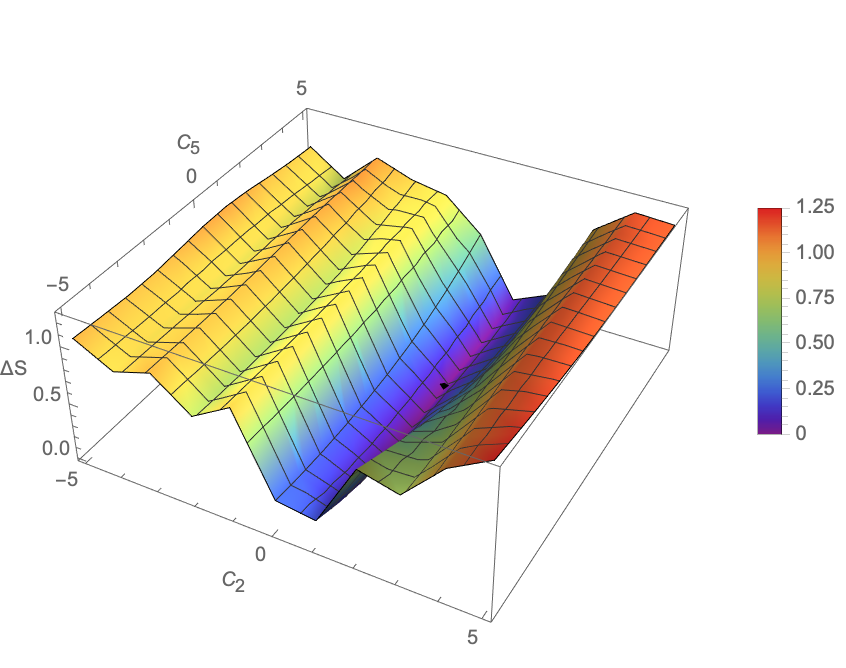}
\includegraphics[width=6cm,height=6cm]{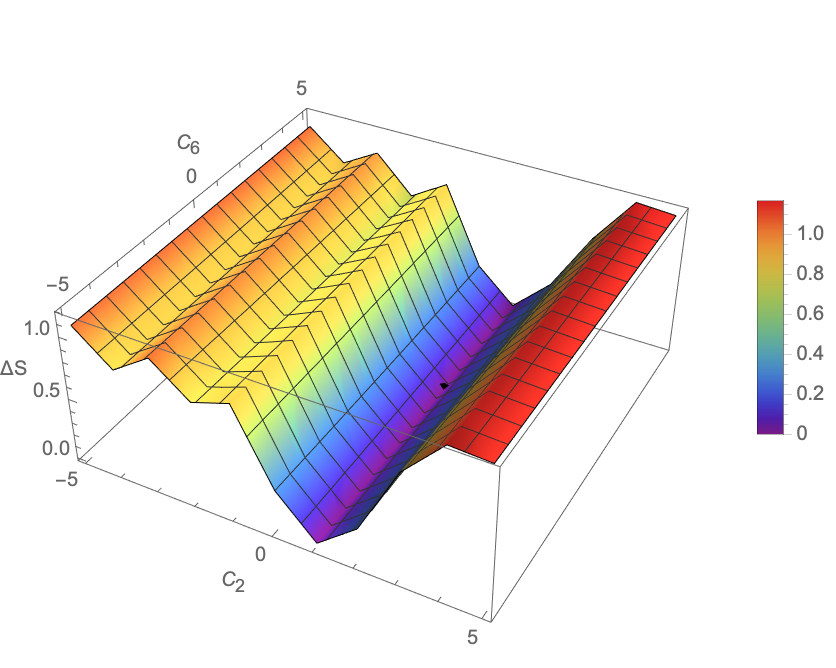}
\includegraphics[width=6cm,height=6cm]{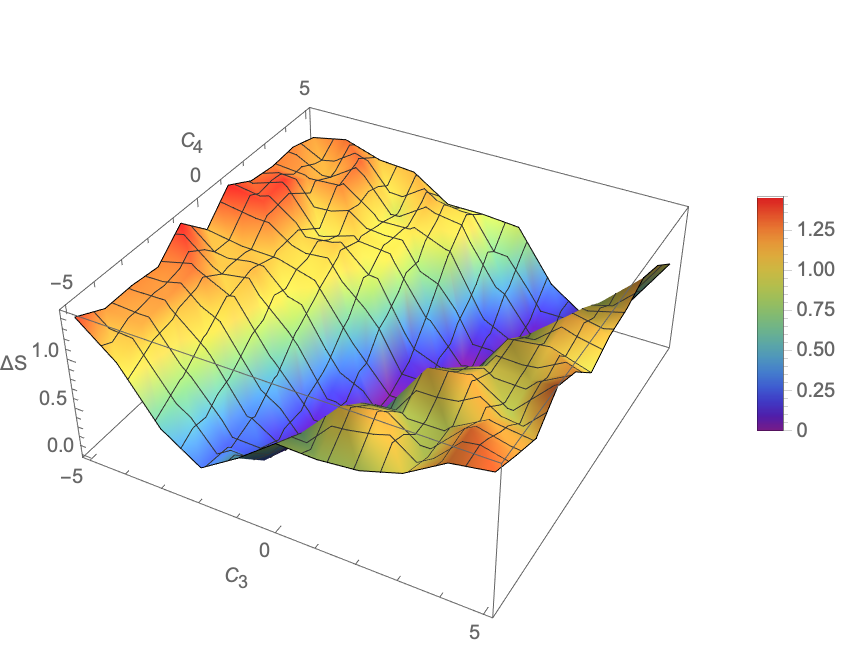}
\includegraphics[width=6cm,height=6cm]{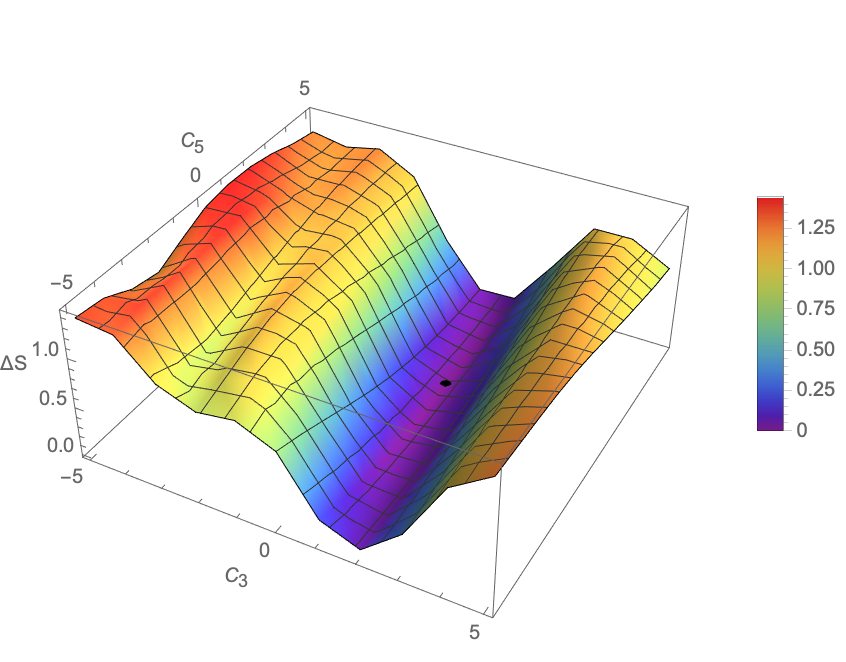}
\includegraphics[width=6cm,height=6cm]{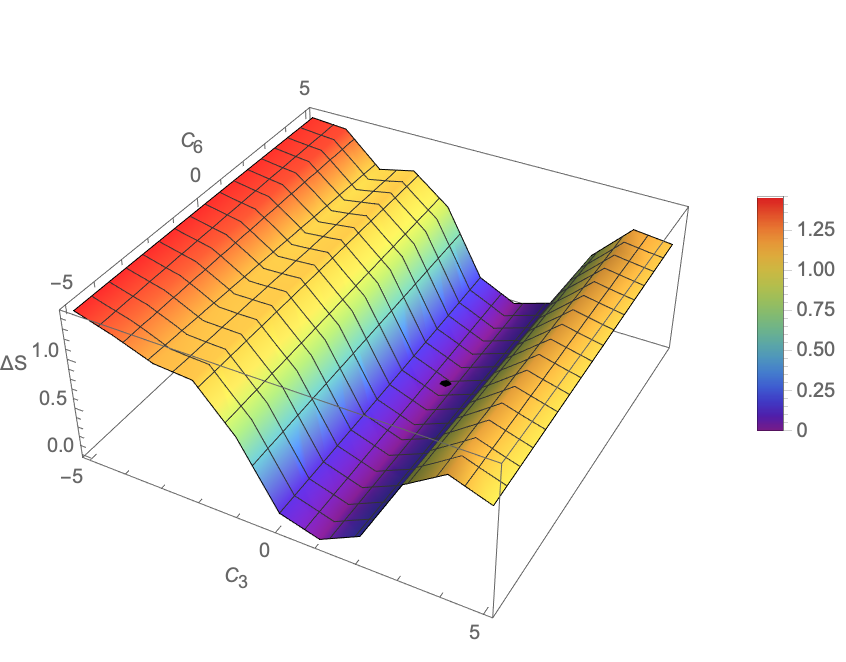}
\includegraphics[width=6cm,height=6cm]{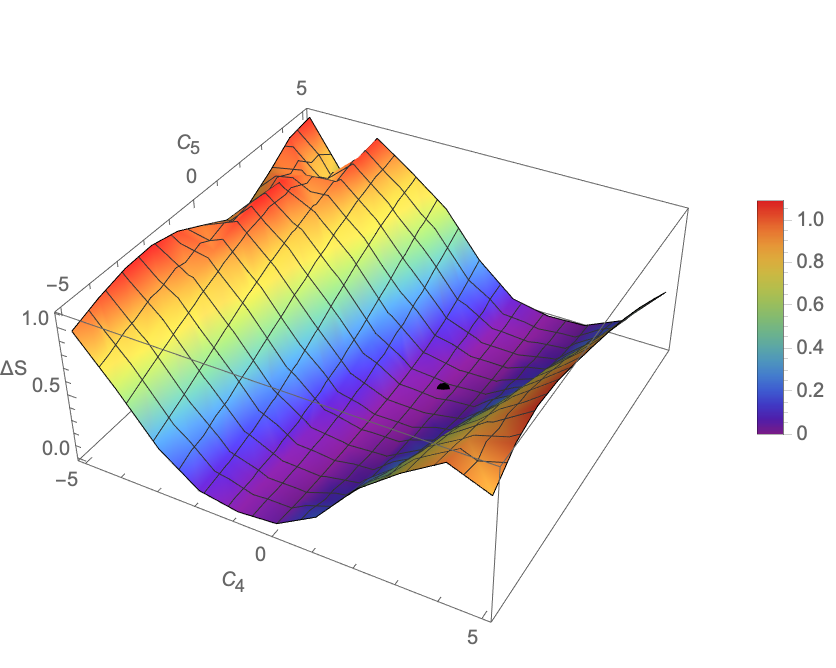}
\includegraphics[width=6cm,height=6cm]{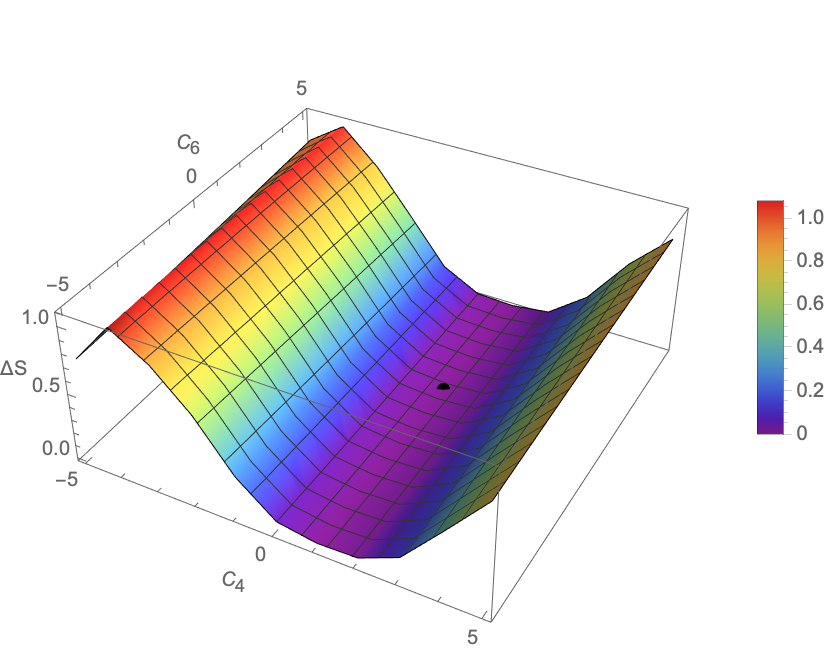}
\includegraphics[width=6cm,height=6cm]{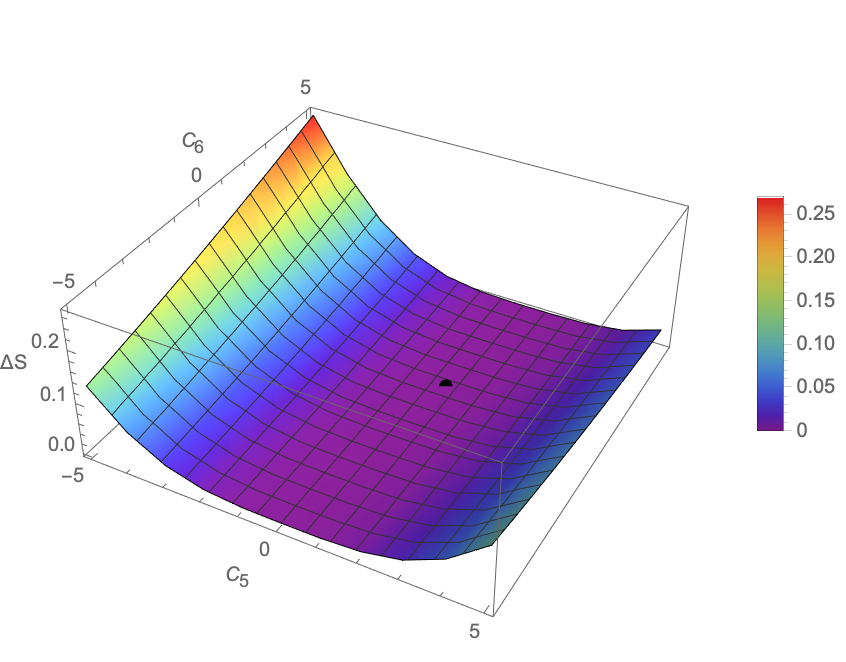}
\endgroup
\caption{Relative Von Neumann entropy $\Delta S$ for massive spin-3 particles. The initial state is set to $\lvert{\rm in}\rangle=\lvert\upuparrows\rangle$ and the kinematic parameters are given by $|\vec{p}_a| = |\vec{p}_b| = |\vec{p}|$, $m_a = m_b = m$, $\vec{b} = (b, 0,0)$, $Gm^2 = 10^{-4}$, $|\vec{p}|b = 1000$, $|\vec{p}|/m = 100$. The minimum, represented by the black point, corresponds to the Wilson coefficient value $(C_{i}, C_{j})=(1,1)$, $\Delta S\approx 1.26\times10^{-8}$.}
\centering
\label{spin3ca242526}
\end{figure}

\newpage
Finally we give the data for Fig 4. The red highlight corresponds to $C_{a,2} = C_{b,2} = 1$, where we can see $\Delta S$ is indeed the minimum.
\begingroup
\allowdisplaybreaks
\begin{longtable}{|c|c|c||c|c|c||c|c|c||c|c|c|}
\hline
$C_{a,2}$ & $C_{b,2}$ & $\Delta S$ & $C_{a,2}$ & $C_{b,2}$ & $\Delta S$ & $C_{a,2}$ & $C_{b,2}$ & $\Delta S$ & $C_{a,2}$ & $C_{b,2}$ & $\Delta S$ \\
\hline
 -5. & -5. & 1.08 & -2.5 & -2.5 & 0.807 & 0 & 0 & 0.282 & 2.5 & 2.5 & 0.607 \\
 -5. & -4.5 & 0.976 & -2.5 & -2. & 0.843 & 0 & 0.5 & 0.102 & 2.5 & 3. & 0.740 \\
 -5. & -4. & 1.09 & -2.5 & -1.5 & 0.747 & 0 & 1. & 0.0000219 & 2.5 & 3.5 & 0.793 \\
 -5. & -3.5 & 0.869 & -2.5 & -1. & 0.820 & 0 & 1.5 & 0.103 & 2.5 & 4. & 0.791 \\
 -5. & -3. & 0.888 & -2.5 & -0.5 & 0.931 & 0 & 2. & 0.282 & 2.5 & 4.5 & 0.851 \\
 -5. & -2.5 & 0.823 & -2.5 & 0 & 0.820 & 0 & 2.5 & 0.462 & 2.5 & 5. & 0.941 \\
 -5. & -2. & 0.989 & -2.5 & 0.5 & 0.469 & 0 & 3. & 0.608 & 3. & -5. & 1.15 \\
 -5. & -1.5 & 0.901 & -2.5 & 1. & 0.000184 & 0 & 3.5 & 0.714 & 3. & -4.5 & 1.08 \\
 -5. & -1. & 0.893 & -2.5 & 1.5 & 0.427 & 0 & 4. & 0.790 & 3. & -4. & 1.01 \\
 -5. & -0.5 & 0.796 & -2.5 & 2. & 0.758 & 0 & 4.5 & 0.835 & 3. & -3.5 & 1.03 \\
 -5. & 0 & 0.802 & -2.5 & 2.5 & 0.815 & 0 & 5. & 0.847 & 3. & -3. & 1.06 \\
 -5. & 0.5 & 0.582 & -2.5 & 3. & 0.992 & 0.5 & -5. & 0.582 & 3. & -2.5 & 0.992 \\
 -5. & 1. & 0.000297 & -2.5 & 3.5 & 1.03 & 0.5 & -4.5 & 0.574 & 3. & -2. & 0.883 \\
 -5. & 1.5 & 0.485 & -2.5 & 4. & 1.03 & 0.5 & -4. & 0.560 & 3. & -1.5 & 0.854 \\
 -5. & 2. & 0.783 & -2.5 & 4.5 & 1.13 & 0.5 & -3.5 & 0.539 & 3. & -1. & 0.877 \\
 -5. & 2.5 & 0.963 & -2.5 & 5. & 1.13 & 0.5 & -3. & 0.510 & 3. & -0.5 & 0.802 \\
 -5. & 3. & 1.15 & -2. & -5. & 0.989 & 0.5 & -2.5 & 0.469 & 3. & 0 & 0.608 \\
 -5. & 3.5 & 1.15 & -2. & -4.5 & 0.972 & 0.5 & -2. & 0.415 & 3. & 0.5 & 0.273 \\
 -5. & 4. & 1.12 & -2. & -4. & 0.905 & 0.5 & -1.5 & 0.347 & 3. & 1. & 0.0000760 \\
 -5. & 4.5 & 1.13 & -2. & -3.5 & 0.820 & 0.5 & -1. & 0.269 & 3. & 1.5 & 0.258 \\
 -5. & 5. & 1.03 & -2. & -3. & 0.830 & 0.5 & -0.5 & 0.185 & 3. & 2. & 0.546 \\
 -4.5 & -5. & 0.976 & -2. & -2.5 & 0.843 & 0.5 & 0 & 0.102 & 3. & 2.5 & 0.740 \\
 -4.5 & -4.5 & 1.08 & -2. & -2. & 0.745 & 0.5 & 0.5 & 0.0338 & 3. & 3. & 0.801 \\
 -4.5 & -4. & 0.922 & -2. & -1.5 & 0.785 & 0.5 & 1. & 6.08 $\times 10^{-6}$ & 3. & 3.5 & 0.835 \\
 -4.5 & -3.5 & 0.842 & -2. & -1. & 0.900 & 0.5 & 1.5 & 0.0339 & 3. & 4. & 0.958 \\
 -4.5 & -3. & 0.782 & -2. & -0.5 & 0.917 & 0.5 & 2. & 0.103 & 3. & 4.5 & 1.08 \\
 -4.5 & -2.5 & 0.955 & -2. & 0 & 0.769 & 0.5 & 2.5 & 0.187 & 3. & 5. & 1.17 \\
 -4.5 & -2. & 0.972 & -2. & 0.5 & 0.415 & 0.5 & 3. & 0.273 & 3.5 & -5. & 1.15 \\
 -4.5 & -1.5 & 0.828 & -2. & 1. & 0.000147 & 0.5 & 3.5 & 0.354 & 3.5 & -4.5 & 1.15 \\
 -4.5 & -1. & 0.901 & -2. & 1.5 & 0.388 & 0.5 & 4. & 0.424 & 3.5 & -4. & 1.16 \\
 -4.5 & -0.5 & 0.782 & -2. & 2. & 0.726 & 0.5 & 4.5 & 0.482 & 3.5 & -3.5 & 1.11 \\
 -4.5 & 0 & 0.833 & -2. & 2.5 & 0.830 & 0.5 & 5. & 0.526 & 3.5 & -3. & 1.02 \\
 -4.5 & 0.5 & 0.574 & -2. & 3. & 0.883 & 1. & -5. & 0.000297 & 3.5 & -2.5 & 1.03 \\
 -4.5 & 1. & 0.000288 & -2. & 3.5 & 1.03 & 1. & -4.5 & 0.000288 & 3.5 & -2. & 1.03 \\
 -4.5 & 1.5 & 0.486 & -2. & 4. & 1.01 & 1. & -4. & 0.000272 & 3.5 & -1.5 & 0.912 \\
 -4.5 & 2. & 0.749 & -2. & 4.5 & 1.01 & 1. & -3.5 & 0.000248 & 3.5 & -1. & 0.875 \\
 -4.5 & 2.5 & 1.00 & -2. & 5. & 1.11 & 1. & -3. & 0.000219 & 3.5 & -0.5 & 0.876 \\
 -4.5 & 3. & 1.08 & -1.5 & -5. & 0.901 & 1. & -2.5 & 0.000184 & 3.5 & 0 & 0.714 \\
 -4.5 & 3.5 & 1.15 & -1.5 & -4.5 & 0.828 & 1. & -2. & 0.000147 & 3.5 & 0.5 & 0.354 \\
 -4.5 & 4. & 1.13 & -1.5 & -4. & 0.839 & 1. & -1.5 & 0.000111 & 3.5 & 1. & 0.000111 \\
 -4.5 & 4.5 & 1.16 & -1.5 & -3.5 & 0.881 & 1. & -1. & 0.0000759 & 3.5 & 1.5 & 0.325 \\
 -4.5 & 5. & 1.09 & -1.5 & -3. & 0.824 & 1. & -0.5 & 0.0000458 & 3.5 & 2. & 0.634 \\
 -4. & -5. & 1.09 & -1.5 & -2.5 & 0.747 & 1. & 0 & 0.0000219 & 3.5 & 2.5 & 0.793 \\
 -4. & -4.5 & 0.922 & -1.5 & -2. & 0.785 & 1. & 0.5 & 6.08 $\times 10^{-6}$ & 3.5 & 3. & 0.835 \\
 -4. & -4. & 0.835 & -1.5 & -1.5 & 0.882 & \textcolor{red}{1.} & \textcolor{red}{1.} & \textcolor{red}{1.26 $\times 10^{-8}$} & 3.5 & 3.5 & 0.993 \\
 -4. & -3.5 & 0.804 & -1.5 & -1. & 0.942 & 1. & 1.5 & 6.08 $\times 10^{-6}$ & 3.5 & 4. & 1.15 \\
 -4. & -3. & 0.908 & -1.5 & -0.5 & 0.874 & 1. & 2. & 0.0000219 & 3.5 & 4.5 & 1.17 \\
 -4. & -2.5 & 1.01 & -1.5 & 0 & 0.701 & 1. & 2.5 & 0.0000458 & 3.5 & 5. & 1.13 \\
 -4. & -2. & 0.905 & -1.5 & 0.5 & 0.347 & 1. & 3. & 0.0000760 & 4. & -5. & 1.12 \\
 -4. & -1.5 & 0.839 & -1.5 & 1. & 0.000111 & 1. & 3.5 & 0.000111 & 4. & -4.5 & 1.13 \\
 -4. & -1. & 0.835 & -1.5 & 1.5 & 0.333 & 1. & 4. & 0.000148 & 4. & -4. & 1.16 \\
 -4. & -0.5 & 0.787 & -1.5 & 2. & 0.665 & 1. & 4.5 & 0.000185 & 4. & -3.5 & 1.15 \\
 -4. & 0 & 0.867 & -1.5 & 2.5 & 0.846 & 1. & 5. & 0.000219 & 4. & -3. & 1.14 \\
 -4. & 0.5 & 0.560 & -1.5 & 3. & 0.854 & 1.5 & -5. & 0.485 & 4. & -2.5 & 1.03 \\
 -4. & 1. & 0.000272 & -1.5 & 3.5 & 0.912 & 1.5 & -4.5 & 0.486 & 4. & -2. & 1.01 \\
 -4. & 1.5 & 0.482 & -1.5 & 4. & 1.01 & 1.5 & -4. & 0.482 & 4. & -1.5 & 1.01 \\
 -4. & 2. & 0.727 & -1.5 & 4.5 & 1.00 & 1.5 & -3.5 & 0.471 & 4. & -1. & 0.901 \\
 -4. & 2.5 & 1.01 & -1.5 & 5. & 0.979 & 1.5 & -3. & 0.453 & 4. & -0.5 & 0.882 \\
 -4. & 3. & 1.01 & -1. & -5. & 0.893 & 1.5 & -2.5 & 0.427 & 4. & 0 & 0.790 \\
 -4. & 3.5 & 1.16 & -1. & -4.5 & 0.901 & 1.5 & -2. & 0.388 & 4. & 0.5 & 0.424 \\
 -4. & 4. & 1.16 & -1. & -4. & 0.835 & 1.5 & -1.5 & 0.333 & 4. & 1. & 0.000148 \\
 -4. & 4.5 & 1.07 & -1. & -3.5 & 0.770 & 1.5 & -1. & 0.264 & 4. & 1.5 & 0.377 \\
 -4. & 5. & 1.17 & -1. & -3. & 0.772 & 1.5 & -0.5 & 0.184 & 4. & 2. & 0.704 \\
 -3.5 & -5. & 0.869 & -1. & -2.5 & 0.820 & 1.5 & 0 & 0.103 & 4. & 2.5 & 0.791 \\
 -3.5 & -4.5 & 0.842 & -1. & -2. & 0.900 & 1.5 & 0.5 & 0.0339 & 4. & 3. & 0.958 \\
 -3.5 & -4. & 0.804 & -1. & -1.5 & 0.942 & 1.5 & 1. & 6.08 $\times 10^{-6}$ & 4. & 3.5 & 1.15 \\
 -3.5 & -3.5 & 0.912 & -1. & -1. & 0.896 & 1.5 & 1.5 & 0.0337 & 4. & 4. & 1.16 \\
 -3.5 & -3. & 1.06 & -1. & -0.5 & 0.794 & 1.5 & 2. & 0.101 & 4. & 4.5 & 1.18 \\
 -3.5 & -2.5 & 0.993 & -1. & 0 & 0.604 & 1.5 & 2.5 & 0.181 & 4. & 5. & 1.18 \\
 -3.5 & -2. & 0.820 & -1. & 0.5 & 0.269 & 1.5 & 3. & 0.258 & 4.5 & -5. & 1.13 \\
 -3.5 & -1.5 & 0.881 & -1. & 1. & 0.0000759 & 1.5 & 3.5 & 0.325 & 4.5 & -4.5 & 1.16 \\
 -3.5 & -1. & 0.770 & -1. & 1.5 & 0.264 & 1.5 & 4. & 0.377 & 4.5 & -4. & 1.07 \\
 -3.5 & -0.5 & 0.824 & -1. & 2. & 0.580 & 1.5 & 4.5 & 0.417 & 4.5 & -3.5 & 1.18 \\
 -3.5 & 0 & 0.875 & -1. & 2.5 & 0.785 & 1.5 & 5. & 0.446 & 4.5 & -3. & 1.14 \\
 -3.5 & 0.5 & 0.539 & -1. & 3. & 0.877 & 2. & -5. & 0.783 & 4.5 & -2.5 & 1.13 \\
 -3.5 & 1. & 0.000248 & -1. & 3.5 & 0.875 & 2. & -4.5 & 0.749 & 4.5 & -2. & 1.01 \\
 -3.5 & 1.5 & 0.471 & -1. & 4. & 0.901 & 2. & -4. & 0.727 & 4.5 & -1.5 & 1.00 \\
 -3.5 & 2. & 0.733 & -1. & 4.5 & 0.954 & 2. & -3.5 & 0.733 & 4.5 & -1. & 0.954 \\
 -3.5 & 2.5 & 0.949 & -1. & 5. & 0.986 & 2. & -3. & 0.754 & 4.5 & -0.5 & 0.889 \\
 -3.5 & 3. & 1.03 & -0.5 & -5. & 0.796 & 2. & -2.5 & 0.758 & 4.5 & 0 & 0.835 \\
 -3.5 & 3.5 & 1.11 & -0.5 & -4.5 & 0.782 & 2. & -2. & 0.726 & 4.5 & 0.5 & 0.482 \\
 -3.5 & 4. & 1.15 & -0.5 & -4. & 0.787 & 2. & -1.5 & 0.665 & 4.5 & 1. & 0.000185 \\
 -3.5 & 4.5 & 1.18 & -0.5 & -3.5 & 0.824 & 2. & -1. & 0.580 & 4.5 & 1.5 & 0.417 \\
 -3.5 & 5. & 1.07 & -0.5 & -3. & 0.877 & 2. & -0.5 & 0.453 & 4.5 & 2. & 0.736 \\
 -3. & -5. & 0.888 & -0.5 & -2.5 & 0.931 & 2. & 0 & 0.282 & 4.5 & 2.5 & 0.851 \\
 -3. & -4.5 & 0.782 & -0.5 & -2. & 0.917 & 2. & 0.5 & 0.103 & 4.5 & 3. & 1.08 \\
 -3. & -4. & 0.908 & -0.5 & -1.5 & 0.874 & 2. & 1. & 0.0000219 & 4.5 & 3.5 & 1.17 \\
 -3. & -3.5 & 1.06 & -0.5 & -1. & 0.794 & 2. & 1.5 & 0.101 & 4.5 & 4. & 1.18 \\
 -3. & -3. & 1.02 & -0.5 & -0.5 & 0.670 & 2. & 2. & 0.272 & 4.5 & 4.5 & 1.19 \\
 -3. & -2.5 & 0.841 & -0.5 & 0 & 0.461 & 2. & 2.5 & 0.429 & 4.5 & 5. & 1.25 \\
 -3. & -2. & 0.830 & -0.5 & 0.5 & 0.185 & 2. & 3. & 0.546 & 5. & -5. & 1.03 \\
 -3. & -1.5 & 0.824 & -0.5 & 1. & 0.0000458 & 2. & 3.5 & 0.634 & 5. & -4.5 & 1.09 \\
 -3. & -1. & 0.772 & -0.5 & 1.5 & 0.184 & 2. & 4. & 0.704 & 5. & -4. & 1.17 \\
 -3. & -0.5 & 0.877 & -0.5 & 2. & 0.453 & 2. & 4.5 & 0.736 & 5. & -3.5 & 1.07 \\
 -3. & 0 & 0.856 & -0.5 & 2.5 & 0.661 & 2. & 5. & 0.737 & 5. & -3. & 1.18 \\
 -3. & 0.5 & 0.510 & -0.5 & 3. & 0.802 & 2.5 & -5. & 0.963 & 5. & -2.5 & 1.13 \\
 -3. & 1. & 0.000219 & -0.5 & 3.5 & 0.876 & 2.5 & -4.5 & 1.00 & 5. & -2. & 1.11 \\
 -3. & 1.5 & 0.453 & -0.5 & 4. & 0.882 & 2.5 & -4. & 1.01 & 5. & -1.5 & 0.979 \\
 -3. & 2. & 0.754 & -0.5 & 4.5 & 0.889 & 2.5 & -3.5 & 0.949 & 5. & -1. & 0.986 \\
 -3. & 2.5 & 0.863 & -0.5 & 5. & 0.899 & 2.5 & -3. & 0.863 & 5. & -0.5 & 0.899 \\
 -3. & 3. & 1.06 & 0 & -5. & 0.802 & 2.5 & -2.5 & 0.815 & 5. & 0 & 0.847 \\
 -3. & 3.5 & 1.02 & 0 & -4.5 & 0.833 & 2.5 & -2. & 0.830 & 5. & 0.5 & 0.526 \\
 -3. & 4. & 1.14 & 0 & -4. & 0.867 & 2.5 & -1.5 & 0.846 & 5. & 1. & 0.000219 \\
 -3. & 4.5 & 1.14 & 0 & -3.5 & 0.875 & 2.5 & -1. & 0.785 & 5. & 1.5 & 0.446 \\
 -3. & 5. & 1.18 & 0 & -3. & 0.856 & 2.5 & -0.5 & 0.661 & 5. & 2. & 0.737 \\
 -2.5 & -5. & 0.823 & 0 & -2.5 & 0.820 & 2.5 & 0 & 0.462 & 5. & 2.5 & 0.941 \\
 -2.5 & -4.5 & 0.955 & 0 & -2. & 0.769 & 2.5 & 0.5 & 0.187 & 5. & 3. & 1.17 \\
 -2.5 & -4. & 1.01 & 0 & -1.5 & 0.701 & 2.5 & 1. & 0.0000458 & 5. & 3.5 & 1.13 \\
 -2.5 & -3.5 & 0.993 & 0 & -1. & 0.604 & 2.5 & 1.5 & 0.181 & 5. & 4. & 1.18 \\
 -2.5 & -3. & 0.841 & 0 & -0.5 & 0.461 & 2.5 & 2. & 0.429 & 5. & 4.5 & 1.25 \\
\hline
\caption{Data set for Fig 4.}
\end{longtable}
\endgroup

\end{document}

%% file: shortcuts.tex
\newcommand*{\la}{\langle}
\newcommand*{\ra}{\rangle}

\newcommand{\AB}[1]{\langle #1 \rangle}
\newcommand{\SB}[1]{[ #1 ]}
\newcommand{\MixLeft}[3]{\langle #1 | #2 | #3 ]}

\newcommand{\BS}[1]{\boldsymbol{ #1 }}

\newcommand{\eq}{\begin{equation}}
\newcommand{\eqe}{\end{equation}}
\newcommand{\eqa}{\begin{eqnarray}}
\newcommand{\eqae}{\end{eqnarray}}